# Modal Phenomena of Surface and Bulk Polaritons in Magnetic-Semiconductor Superlattices

Vladimir R. Tuz,[1,2] Illia V. Fedorin,[3] Volodymyr I. Fesenko[1,2,*]

[1]*International Center of Future Science, State Key Laboratory on Integrated Optoelectronics, College of Electronic Science and Engineering, Jilin University, Changchun 130012, China*
[2]*Institute of Radio Astronomy of National Academy of Sciences of Ukraine, Kharkiv 61002, Ukraine*
[3]*Department of Materials for Electronics and Solar Cells, National Technical University `Kharkiv Polytechnical Institute', Kharkiv 61002, Ukraine*
*\*Corresponding author: [volodymyr.i.fesenko@gmail.com](mailto:volodymyr.i.fesenko@gmail.com)*

**We discuss peculiarities of bulk and surface polaritons propagating in a composite magnetic-semiconductor superlattice influenced by an external static magnetic field. Three particular configurations of magnetization, namely the Voigt, polar and Faraday geometries are considered. In the long-wavelength limit, involving the effective medium theory, the proposed superlattice is described as an anisotropic uniform medium defined by the tensors of effective permittivity and effective permeability. The study is carried out in the frequency band where the characteristic resonant frequencies of the underlying constitutive magnetic and semiconductor materials of the superlattice are different but closely spaced. The effects of mode crossing and anti-crossing in dispersion characteristics of both bulk and surface polaritons are revealed and explained with an assistance of the concept of Morse critical points from the catastrophe theory.** © 2017 Optical Society of America

## 1. INTRODUCTION

Surface polaritons are a special type of electromagnetic waves propagating along the interface of two partnering materials whose material functions (e.g., permittivities) have opposite signs that is typical for a metal-dielectric boundary [1]. These waves are strongly localized at the interface and penetrate into the surrounding space over a distance of wavelength order in a medium, and their amplitudes fall exponentially away from the surface. Observed strong confinement of electromagnetic field in small volumes beyond the diffraction limit, leads to enormous increasing matter-field interaction and it makes attractive using surface waves in the wide fields from microwave and photonic devices to solar cells [2], [3]. Furthermore, surface waves are highly promising in view of material physics because from the character of waves propagation one can derive information about both interface quality and electromagnetic properties of partnering materials (such as permittivity and permeability) [1]. High sensitivity to the electromagnetic properties of media enables utilization of surface waves in sensing applications, particularly in both chemical and biological systems [4]. Thus, studying characteristics of surface waves is essential in the physics of surfaces and optics; in the latter case, the research has led to the emergence of a new field – plasmonics.

Today plasmonics is rapidly developing science characterized by enormous variety of possible practical applications. In many of them, an ability to control and guide surface waves is a crucial characteristic. Thereby in last decades many efforts have been made in order to realize active tunable components for plasmonic integrated circuits such as switchers, active couplers, modulators, etc. In this regard, searching efficient ways to control characteristics of plasmon-polariton propagation by utilizing external driving agents is a very important task. In particular, the nonlinear, thermo-optical and electro-optical effects are proposed to be used in the tunable plasmonic devices for the control of plasmon-polariton propagation [5]–[8]. In such devices the tuning mechanism is conditioned by changing the permittivity of the dielectric medium due to applying an external electric field or temperature control. At the same time, utilization of the external magnetic field as a driving agent to gain a control over polaritons dispersion features is very promising, since it allows changing both the permeability of magnetic materials (e.g., ferrites) and permittivity of conducting materials (e.g., metals and semiconductors). It is worth mentioning that the uniqueness of this controlling mechanism lies in the fact that the polaritons properties depend not only on the magnitude of the magnetic field, but also on its direction. An applied magnetic field also produces additional branches in spectra of surface magnetic plasmon-polariton resulting in the multiband propagation accompanied by nonreciprocal effects [9]–[16]. Thus, combination of plasmonic and magnetic functionalities opens a prospect towards active devices with an additional degree of freedom in the control of plasmon-polariton properties, and such systems have already found a number of practical applications in integrated photonic devices for telecommunications (see, for instance, [3], [8] and references therein).

In this framework, using superlattices (which typically consist of alternating layers of two partnering materials) that are capable to provide a combined plasmon and magnetic functionality instead of traditional plasmonic systems (in which the presence of a metal-dielectric interface is implied) has great prospects. Particularly it conditioned by the fact that the superlattices demonstrate many exotic electronic and optical properties uncommon to the homogeneous (bulk) samples due to the presence of an additional periodic potential whose period is greater than the original lattice constant [14]. The applying of an external magnetic field to a superlattice results in arising so-called magnetoplasmon-polariton excitations. Properties of the magnetic polaritons in superlattices of different kind being under the action of an external static magnetic field have been studied by many authors during several last decades [10]–[16]. The problem is usually solved within two distinct considerations of gyroelectric media (e.g., semiconductors) with magnetoplasmons [10], [14] and gyromagnetic media (e.g., ferrites) with magnons [11]-[13], [16] which involve the medium characterization with either permittivity or permeability tensor having asymmetric off-diagonal parts. This distinction is governed by the fact that the resonant frequencies of permeability of magnetic materials usually lie in the microwave range whereas characteristic frequencies of permittivity of semiconductors commonly are in the infrared range.

At the same time, it is evident that combining together magnetic and semiconductor materials into a single gyroelectromagnetic superlattice in which both permeability and permittivity simultaneously are tensor quantities allows additional possibilities in the control of polaritons using the magnetic field, that are unattainable in convenient either gyromagnetic or gyroelectric media. Fortunately, it is possible to design heterostructures in which both characteristic resonant frequencies of semiconductor and magnetic materials can be different but, nevertheless, closely spaced in the same frequency band. As a relevant example the magnetic-semiconductor heterostructures proposed in [17]-[20] can be mentioned that are able to exhibit a gyroelectromagnetic effect from gigahertz up to tens of terahertz [21]. Thus investigation of the electromagnetic properties of such structures in view of their promising application as a part of plasmonic devices is a significant task.

This paper is devoted to the study of crossing and anti-crossing effects and other dispersion peculiarities of both bulk and surface polaritons propagating in a finely-stratified magnetic-semiconductor superlattice influenced by an external static magnetic field. It is organized as follows. In Section 2 we describe the superlattice under study and formulate the problem to be solved. Section 3 represents the problem solution in a general form assuming an arbitrary orientation of the external magnetic field with respect to the direction of wave propagation and interface of the structure. In Section 4 we reveal dispersion peculiarities of bulk and surface polaritons in the given superlattice for three particular cases of the vector orientation of the external magnetic field with respect to the superlattice's interface and wavevector, namely we study the configurations where the external magnetic field is influenced in the Voigt, polar and Faraday geometries. Finally, Section 5 summarizes the paper. Appendix A gives expressions of the effective medium theory used here for calculation of the dispersion characteristics of modes under the long-wavelength approximation. A brief introduction to the concept of the Morse critical points from the catastrophe theory which is applied to description of mode crossing and anti-crossing effects is presented in Appendix B.

## 2. OUTLINE OF PROBLEM

Thereby, in this paper we study dispersion features of surface and bulk polaritons propagating in a *semi-infinite* stack of identical composite double-layered slabs arranged along the $y$-axis that forms a *superlattice* (Fig. 1). Each composite slab within the superlattice includes magnetic (with constitutive parameters $\varepsilon_m$, $\hat{\mu}_m$) and semiconductor (with constitutive parameters $\hat{\varepsilon}_s$, $\mu_s$) layers having thicknesses $d_m$ and $d_s$, respectively. The stack possesses a periodic structure (with period $L = d_m + d_s$) that fills half-space $y < 0$ and adjoins a vacuum ($\varepsilon_0 = \mu_0 = 1$) occupying half-space $y > 0$. Therefore, the superlattice's interface lies in the $x - z$ plane, and along this plane the system is considered to be infinite. The structure under investigation is influenced by an external static magnetic field $\vec{M}$ that lies in the $y - z$ plane and makes an angle $\theta$ with the $y$-axis. It is supposed that the strength of this field is high enough to form a homogeneous saturated state of magnetic as well as semiconductor subsystems. Finally, the wavevector $\vec{k}$ of the macroscopic electric field lies in the $x - z$ plane and makes an angle $\varphi$ with the $x$-axis.

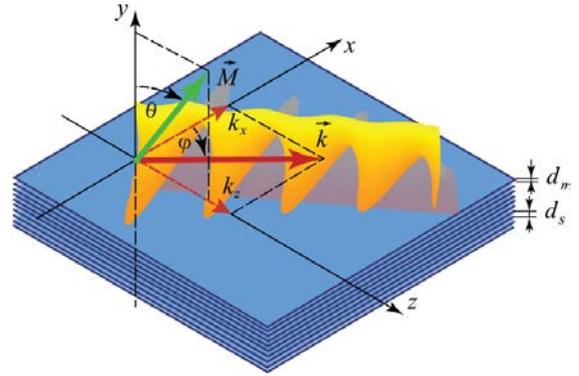

**Fig. 1.** Schematic representation of a magnetic-semiconductor superlattice which is influenced by an external static magnetic field $\vec{M}$, and a visual representation of the tangential electric field distribution of the surface polariton propagating over the interface between the given structure and free space.

In the general case, when no restrictions are imposed on characteristic dimensions ($d_m$, $d_s$ and $L$) of the superlattice compared to the wavelength of propagating modes, the transfer matrix formalism [22, 23] is usually involved in order to reveal the dispersion features of polaritons. It implies a *numerical* solution of a canonical boundary-value problem formulated for each layer within the period of superlattice, and then performing a subsequent multiplication of the obtained transfer matrices to form a semi-infinite extent. On the other hand, when all characteristic dimensions of the superlattice satisfy the long-wavelength limit, i.e. they are all much smaller than the wavelength in the corresponding layer and period of structure ($d_m \ll \lambda, d_s \ll \lambda, L \ll \lambda$), homogenization procedures from the effective medium theory can be involved in order to derive averaged expressions for effective constitutive parameters of the superlattice in an *explicit* form (see, Refs. [24]-[29] and Appendix A), that is suitable for identifying the main features of interest. Therefore further only the modes under the long-wavelength approximation are studied in this paper, i.e. the structure is considered to be a *finely-stratified* one.

For further reference, the dispersion curves of the tensor components of relative effective permeability $\hat{\mu}_{eff}$ and relative effective permittivity $\hat{\varepsilon}_{eff}$ (see, Eq. (A7) in Appendix A) of the homogenized medium (with filling factors $\delta_m = \delta_s = 0.5$) are presented in Fig. 2. In this figure panels (a) and (b) represent constitutive parameters for the polar configuration, whereas panels (c) and (d) represent those for the Voigt and Faraday configurations of magnetization. For these calculations we used typical constitutive parameters for magnetic and semiconductor materials. In particular, here we follow the results of paper [30], where a magnetic-semiconductor composite is made in the form of a barium-cobalt/doped-silicon superlattice for operating in the microwave part of spectrum. A distinct peculiarity of such a superlattice is that the characteristic resonant frequencies of the underlying constitutive magnetic and semiconductor materials are different but rather closely spaced within the same frequency band.

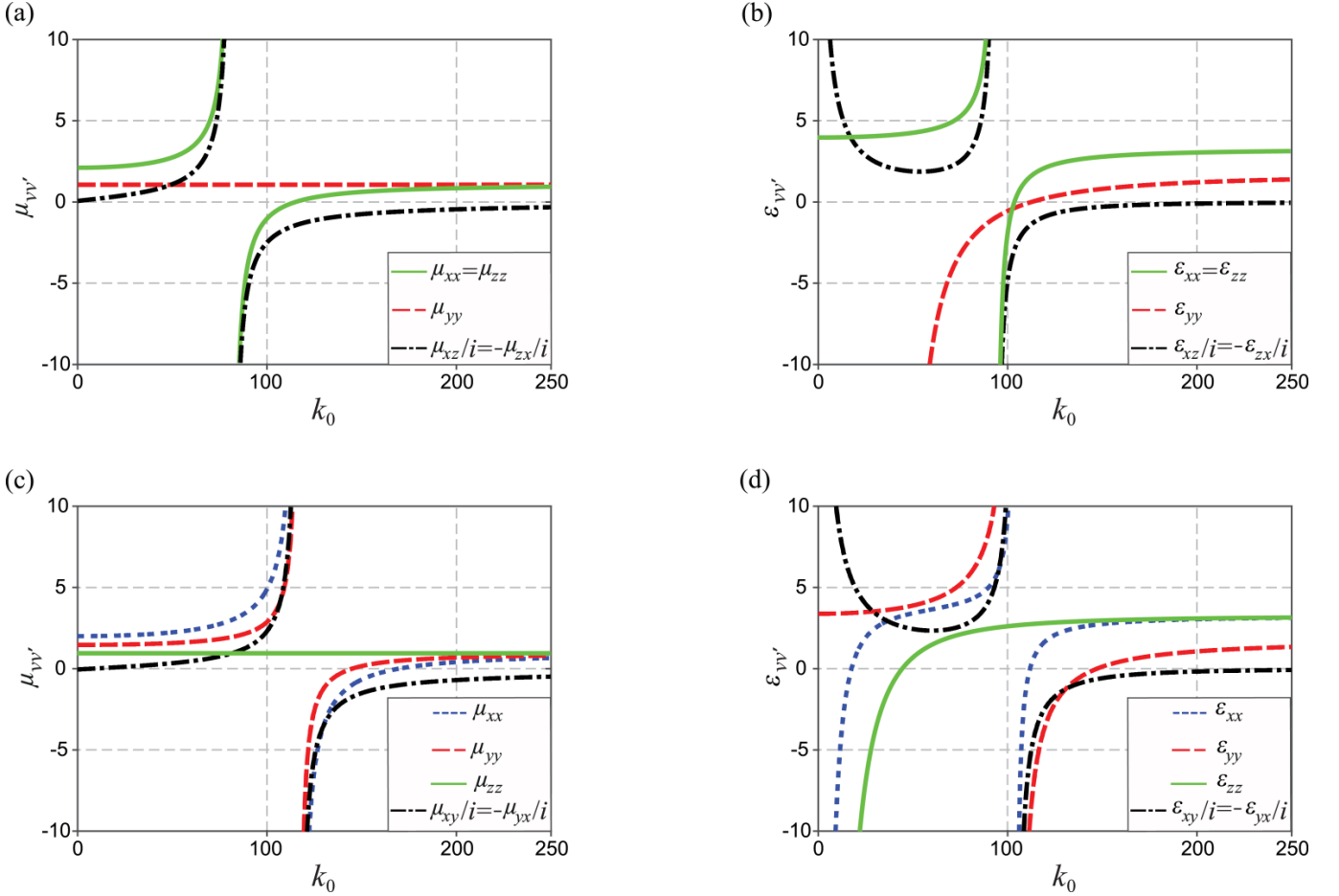

**Fig. 2.** Dispersion curves of tensors components of (a, c) relative effective permeability $\hat{\mu}_{eff}$ and (b, d) relative effective permittivity $\hat{\varepsilon}_{eff}$ of the homogenized medium. Panels (a) and (b) correspond to the polar geometry and panels (d) and (c) correspond to the Voigt and Faraday geometries. For the magnetic constitutive layers, under saturation magnetization of 2930 G, parameters are: $f_0 = \omega_0/2\pi = 3.9$ GHz, $f_m = \omega_m/2\pi = 8.2$ GHz, $b = 0$, $\varepsilon_m = 5.5$; for the semiconductor constitutive layers, parameters are: $f_p = \omega_p/2\pi = 5.5$ GHz, $f_c = \omega_c/2\pi = 4.5$ GHz, $\nu = 0$, $\varepsilon_l = 1.0$, $\mu_s = 1.0$. Filling factors are: $\delta_m = \delta_s = 0.5$.

From Fig. 2 one can conclude that in both the Voigt and Faraday geometries the next relations between the components of effective tensor (A7) hold: $\tilde{g}_{xx} \neq \tilde{g}_{yy} \neq \tilde{g}_{zz}$ and $\tilde{g}_{xy} = -\tilde{g}_{yx} \neq 0$, so it means that the obtained homogenized medium is a biaxial bigyrotropic crystal [1]. In the polar geometry it is a uniaxial bigyrotropic crystal, and the following relations between tensors components are met: $\tilde{g}_{xx} = \tilde{g}_{zz} \neq \tilde{g}_{yy}$ and $\tilde{g}_{xz} = -\tilde{g}_{zx} \neq 0$.

To sum up, with an assistance of the homogenization procedures from the effective media theory, the superlattice under study is approximately represented as a uniaxial or biaxial anisotropic uniform medium when an external static magnetic field $\vec{M}$ is directed along or orthogonal to the structure periodicity, respectively. In the latter case, the first optical axis of the biaxial medium is directed along the structure periodicity, whereas the second one coincides with the direction of the external static magnetic field $\vec{M}$.

## 2. GENERAL SOLUTION FOR BULK AND SURFACE POLARITONS

In order to obtain a *general* solution for both bulk and surface polaritons we follow the approach developed in Ref. [10] where dispersion characteristics of polaritons in a uniaxial anisotropic dielectric medium has been studied. Here we extend this approach to the case of a gyroelectromagnetic medium whose permittivity as well as permeability simultaneously are tensor quantities.

In a general form [26], the electric and magnetic field vectors $\vec{E}$ and $\vec{H}$ used here are represented as

$$\vec{P}^{(j)} = \vec{p}^{(j)} \exp[i(k_x x + k_z z)] \exp(\mp \kappa y), \quad (1)$$

where a time factor $\exp(-i\omega t)$ is also supposed and omitted, and sign " − " is related to the fields in the upper medium ($y > 0$, $j = 0$), while sign " + " is related to the fields in the composite medium ($y < 0$, $j = 1$), respectively, which provide required wave attenuation along the $y$-axis.

From a pair of the curl Maxwell's equations $\nabla \times \vec{E} = ik_0\vec{B}$ and $\nabla \times \vec{H} = -ik_0\vec{D}$, in a standard way we arrive at the following equation for the macroscopic field:

$$\nabla \times \nabla \times \vec{P}^{(j)} - k_0^2 \hat{\varsigma}^{(j)} \vec{P}^{(j)} = 0, \quad (2)$$

where $k_0 = \omega/c$ is the free space wavenumber, and $\hat{\varsigma}^{(j)}$ is introduced as the product of $\hat{\mu}^{(j)}$ and $\hat{\varepsilon}^{(j)}$ made in the appropriate order.

For the upper medium ($j = 0$), direct substitution of expression (1) with $\vec{P}^{(0)}$ and corresponding constitutive parameters ($\hat{\varsigma}^{(0)}_{\nu\nu'} = 1$ for $\nu = \nu'$, and $\hat{\varsigma}^{(0)}_{\nu\nu'} = 0$ for $\nu \neq \nu'$; here and further subscripts $\nu$ and $\nu'$ are substituted to iterate over indexes of the tensors components $x$, $y$ and $z$ in Cartesian coordinates) into Eq. (2) gives us the relation with respect to $\kappa_0$:

$$\kappa_0^2 = k^2 - k_0^2, \quad (3)$$

where $k^2 = k_x^2 + k_z^2$.

For the composite medium ($j = 1$), substitution of (1) with $\vec{P}^{(1)}$ and $\hat{\varsigma}^{(1)}$ into (2) and subsequent elimination of $P_y^{(1)}$, yield us the following system of two homogeneous algebraic equations for the rest components of $\vec{P}^{(1)}$:

$$\begin{aligned} A_{xz}P_x^{(1)} + B_{xz}P_z^{(1)} &= 0, \\ B_{zx}P_x^{(1)} + A_{zx}P_z^{(1)} &= 0, \end{aligned} \quad (4)$$

where $A_{nm}$ and $B_{nm}$ are function of $\kappa$ derived in the form [10]:

$$\begin{aligned} A_{nm}(\kappa) &= \left(k_m^2 - k_0^2\varsigma_{nn}^{(1)} - \kappa^2\right)\varkappa_y^2 + k_n^2\kappa^2 \\ &\quad + ik_n\kappa\left(\varsigma_{ny}^{(1)} + \varsigma_{yn}^{(1)}\right)k_0^2 - k_0^4\varsigma_{ny}^{(1)}\varsigma_{yn}^{(1)}, \\ B_{nm}(\kappa) &= -\left(k_nk_m + \varsigma_{nm}^{(1)}\right)\varkappa_y^2 + k_nk_m\kappa^2 \\ &\quad + i\kappa\left(k_n\varsigma_{ym}^{(1)} + k_m\varsigma_{ny}^{(1)}\right)k_0^2 - k_0^4\varsigma_{ny}^{(1)}\varsigma_{ym}^{(1)}, \end{aligned} \quad (5)$$

with $\varkappa_y^2 = k^2 - k_0^2\varsigma_{yy}^{(1)}$, and subscripts $m$ and $n$ iterate over indexes $x$ and $z$.

In order to find a nontrivial solution of system (5), we set its determinant of coefficients to zero. After disclosure of the determinant, we obtain an equation of the fourth degree with respect to $\kappa$:

$$\varsigma_{yy}^{(1)}\kappa^4 + a\kappa^3 + b\kappa^2 + c\kappa + d = 0, \quad (6)$$

whose coefficients $a, b, c,$ and $d$ are:

$$\begin{aligned} a &= ik_x\left(\varsigma_{yz}^{(1)} + \varsigma_{zy}^{(1)}\right) + ik_z\left(\varsigma_{yx}^{(1)} + \varsigma_{xy}^{(1)}\right), \\ b &= k_0^2\left[\varsigma_{yy}^{(1)}\left(\varsigma_{zz}^{(1)} + \varsigma_{xx}^{(1)}\right) - \varsigma_{zy}^{(1)}\varsigma_{yz}^{(1)} - \varsigma_{xy}^{(1)}\varsigma_{yx}^{(1)}\right] \\ &\quad -\left[k^2\varsigma_{yy}^{(1)} + k_z^2\varsigma_{zz}^{(1)} + k_x^2\varsigma_{xx}^{(1)} + k_xk_z\left(\varsigma_{zx}^{(1)} + \varsigma_{xz}^{(1)}\right)\right], \\ c &= -ik_0^2\left\{k_z\left[\varsigma_{xy}^{(1)}\varsigma_{zx}^{(1)} + \varsigma_{yx}^{(1)}\varsigma_{xz}^{(1)} - \varsigma_{xx}^{(1)}\left(\varsigma_{yz}^{(1)} + \varsigma_{zy}^{(1)}\right)\right]\right. \\ &\quad \left. + k_x\left[\varsigma_{yz}^{(1)}\varsigma_{zx}^{(1)} + \varsigma_{zy}^{(1)}\varsigma_{xz}^{(1)} - \varsigma_{zz}^{(1)}\left(\varsigma_{yx}^{(1)} + \varsigma_{xy}^{(1)}\right)\right]\right\} \\ &\quad -ik^2\left[k_z\left(\varsigma_{yz}^{(1)} + \varsigma_{zy}^{(1)}\right) + k_x\left(\varsigma_{yx}^{(1)} + \varsigma_{xy}^{(1)}\right)\right], \end{aligned}$$

$$\begin{aligned} d &= k^2\left[k_z^2\varsigma_{zz}^{(1)} + k_x^2\varsigma_{xx}^{(1)} + k_xk_z\left(\varsigma_{zx}^{(1)} + \varsigma_{xz}^{(1)}\right)\right] \\ &\quad + k_0^2k^2\left[\varsigma_{zx}^{(1)}\varsigma_{xz}^{(1)} - \varsigma_{zz}^{(1)}\varsigma_{xx}^{(1)}\right] \\ &\quad + k_0^2\left\{k_x^2\left(\varsigma_{xy}^{(1)}\varsigma_{yx}^{(1)} - \varsigma_{yy}^{(1)}\varsigma_{xx}^{(1)}\right) + k_z^2\left(\varsigma_{zy}^{(1)}\varsigma_{yz}^{(1)} - \varsigma_{yy}^{(1)}\varsigma_{zz}^{(1)}\right)\right. \\ &\quad \left. + k_xk_z\left[\varsigma_{xy}^{(1)}\varsigma_{yz}^{(1)} + \varsigma_{zy}^{(1)}\varsigma_{yx}^{(1)} - \varsigma_{yy}^{(1)}\left(\varsigma_{zx}^{(1)} + \varsigma_{xz}^{(1)}\right)\right]\right\} \\ &\quad + k_0^4\left[\varsigma_{xx}^{(1)}\varsigma_{yy}^{(1)}\varsigma_{zz}^{(1)} - \varsigma_{yy}^{(1)}\varsigma_{zx}^{(1)}\varsigma_{xz}^{(1)}\right. \\ &\quad \left. + \varsigma_{xy}^{(1)}\left(\varsigma_{yz}^{(1)}\varsigma_{zx}^{(1)} - \varsigma_{zz}^{(1)}\varsigma_{yx}^{(1)}\right) + \varsigma_{zy}^{(1)}\left(\varsigma_{xz}^{(1)}\varsigma_{yx}^{(1)} - \varsigma_{yz}^{(1)}\varsigma_{xx}^{(1)}\right)\right]. \end{aligned} \quad (7)$$

The dispersion relation for *bulk* polaritons is then obtained straightforwardly from (6) by putting $\kappa = 0$ inside it.

In order to find the dispersion law of surface polaritons from four roots of (6) two physically correct ones must be selected. In general, two such roots are required to satisfy the electromagnetic boundary conditions at the surface of the composite medium. We define these roots as $\kappa_1$ and $\kappa_2$, and then following [10] introduce the quantities $K_w$ ($w = 1,2$) in the form:

$$\begin{aligned} P_x^{(1)}(\kappa_w) &= K_w A_{zx}(\kappa_w), \\ P_y^{(1)}(\kappa_w) &= K_w C(\kappa_w), \\ P_z^{(1)}(\kappa_w) &= -K_w B_{zx}(\kappa_w), \end{aligned} \quad (8)$$

where $C(\kappa_w) = -(1/\varkappa_y^2)\left[\left(ik_x\kappa_w - k_0^2\varsigma_{yx}^{(1)}\right)A_{zx}(\kappa_w) + \left(ik_z\kappa_w - k_0^2\varsigma_{yz}^{(1)}\right)B_{zx}(\kappa_w)\right]$.

In (8), unknown quantities $K_w$ need to be determined from the boundary conditions.

Taking into consideration that two appropriate roots $\kappa_1$ and $\kappa_2$ of (6) are selected, the components of the field $\vec{P}^{(1)}$ can be rewritten as the linear superposition of two terms

$$\begin{aligned} P_x^{(1)} &= \sum_{w=1,2} K_w A_{zx}(\kappa_w) \exp(\kappa_w y), \\ P_y^{(1)} &= \sum_{w=1,2} K_w C(\kappa_w) \exp(\kappa_w y), \\ P_z^{(1)} &= \sum_{w=1,2} K_w B_{zx}(\kappa_w) \exp(\kappa_w y), \end{aligned} \quad (9)$$

where $y < 0$ and the factor $\exp[i(k_x x + k_z z - \omega t)]$ is omitted.

Involving a pair of the divergent Maxwell's equations $\nabla \cdot \vec{B} = 0$ and $\nabla \cdot \vec{D} = 0$ in the form

$$\nabla \cdot \vec{Q}^{(j)} = \nabla \cdot \left(\hat{g}^{(j)} P^{(j)}\right) = 0, \quad (10)$$

where $g$ is substituted for permeability $\mu$ and permittivity $\varepsilon$, and $\vec{Q}$ is substituted for the magnetic $\vec{B}$ and electric $\vec{D}$ flux densities, one can immediately obtain the relations between the field components in the upper ($y > 0$) medium as follow

$$Q_y^{(0)} = (ig_0/\kappa_0)\left(k_x P_x^{(0)} + k_z P_z^{(0)}\right). \quad (11)$$

The boundary conditions at the interface require the continuity of the tangential components of $\vec{E}$ and $\vec{H}$ as well as the normal components of $\vec{D}$ and $\vec{B}$, i.e. in our notations these components are $P_x$, $P_z$ and $Q_y$, respectively. Thus, application of the boundary conditions together with (11) gives us the next set of equations

$$\begin{aligned} (ig_0/\kappa_0)\left(k_x P_x^{(0)} + k_z P_z^{(0)}\right) &= \tilde{g}_{yx} \sum_{w=1,2} K_w A_{zx}(\kappa_w) \\ &\quad + \tilde{g}_{yx} \sum_{w=1,2} K_w C(\kappa_w) - \tilde{g}_{yz} \sum_{w=1,2} K_w B_{zx}(\kappa_w), \\ P_x^{(0)} &= \sum_{w=1,2} K_w A_{zx}(\kappa_w), \\ P_z^{(0)} &= -\sum_{w=1,2} K_w B_{zx}(\kappa_w), \\ k_z P_x^{(0)} - k_x P_z^{(0)} &= k_z \sum_{w=1,2} K_w A_{zx}(\kappa_w) \\ &\quad + k_x \sum_{w=1,2} K_w B_{zx}(\kappa_w). \end{aligned} \quad (12)$$

The system of equations (12) has a nontrivial solution only if its determinant vanishes. Applying this condition gives us the required dispersion equation for *surface* polaritons.

Finally, the amplitudes $K_1$ and $K_2$ can be found by solving set of linear homogeneous equations (12). They are:

$$K_1 = [k_x A_{zx}(\kappa_2) + k_z B_{zx}(\kappa_2)](\kappa_0 + \kappa_2),$$
$$K_2 = -[k_x B_{zx}(\kappa_1) + k_z A_{zx}(\kappa_1)](\kappa_0 + \kappa_1).$$
**(13)**

Here the problem is considered to be formally solved, and the dispersion relations are derived in a general form for both bulk and surface polaritons.

## 3. DISPERSION FEATURES OF BULK AND SURFACE POLARITONS FOR PARTICULAR ORIENTATIONS OF MAGNETIZATION

Since further our goal is to elucidate the dispersion laws of the bulk and surface polaritons (which are in fact *eigenwaves*), we are interested in real solutions of Eq. (6). In order to find the real solutions, absence of losses in constitutive parameters of the underlying layers is supposed.

We consider three particular orientations of the external magnetic field $\vec{M}$ with respect to the superlattice's interface (the $x - z$ plane) and to the wavevector $\vec{k}$, namely: (i) the *polar geometry* in which the external magnetic field is applied perpendicular to both the direction of wave propagation ($\vec{M} \perp \vec{k}$) and structure's interface ($\vec{M} \parallel y$) as shown in Fig. 3(a); (ii) the *Voigt geometry* in which the external magnetic field is applied parallel to the structure's interface and it is perpendicular to the direction of the wave propagation, so $\vec{M} \parallel z$ and $\vec{M} \perp \vec{k}$ as presented in Fig. 3(b); (iii) the *Faraday geometry* in which the external magnetic field is applied parallel to both the direction of wave propagation and structure's interface, i.e. $\vec{M} \parallel z$ and $\vec{M} \parallel \vec{k}$ as presented in Fig. 3(c).

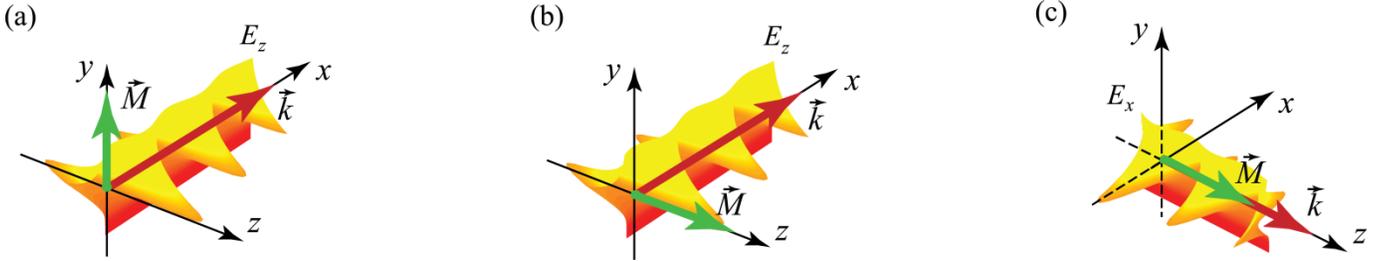

**Fig. 3.** Three particular orientations of the external magnetic field vector $\vec{M}$ with respect to the superlattice's interface and wavevector; (a) polar geometry, $\vec{M} \parallel y$, $\vec{M} \perp \vec{k}$; (b) Voigt geometry, $\vec{M} \parallel z$, $\vec{M} \perp \vec{k}$; (c) Faraday geometry $\vec{M} \parallel z$, $\vec{M} \parallel \vec{k}$.

With respect to the problem of polaritons, in any kind of an unbounded gyrotropic medium there are two distinct eigenwaves (the bulk waves), whereas the surface waves split apart only for some particular configurations (e.g., for the Voigt geometry), and generally, the field has all six field components. Such surface waves are classified as hybrid EH-modes and HE-modes, and these modes appear as some superposition of longitudinal and transverse waves. By analogy with [31] we classify hybrid modes depending on the magnitudes ratio between the longitudinal electric and magnetic fields components ($P_x$ and $P_z$ components for the polar and Faraday geometries, respectively). For instance, for the polar geometry, it is supposed that the wave has the EH-type if $E_x > H_x$ and the HE-type if $H_x > E_x$. Contrariwise, in the Faraday geometry we stipulate that the wave has the EH-type if $E_z > H_z$ and the HE-type if $H_z > E_z$. In the Voigt geometry the waves appears as transverse electric (TE) and transverse magnetic (TM) modes, where each of them has three field components.

The polaritons dispersion relations for these specific cases of magnetization can be obtained by using general results derived in Section 2 with applying appropriate boundary and initial conditions. Further in this study we are mainly interested in a manifestation of the crossing and anti-crossing effects which are considered involving a unified technique about the Morse critical points from the catastrophe theory (for details, see [35]-[40] and Appendix B).

### A. Voigt geometry

When an external static magnetic field is influenced in the Voigt geometry ($\vec{M} \parallel z$, $\vec{M} \perp \vec{k}$), solution of Eq. (6) for both bulk and surface waves splits apart into two independent equations for distinct polarizations [10], namely TE-modes with field components $\{H_x, H_y, E_z\}$ and TM-modes with field components $\{E_x, E_y, H_z\}$. Therefore, regions of existence of bulk polaritons are uniquely determined by solutions of two separated equations related to the TE and the TM modes as follows [26], [32]

$$k_x^2 - k_0^2 \varepsilon_{zz} \mu_v \mu_{yy} \mu_{xx}^{-1} = 0,$$ **(14)**

$$k_x^2 - k_0^2 \mu_{zz} \varepsilon_v \varepsilon_{yy} \varepsilon_{xx}^{-1} = 0,$$ **(15)**

where $\varepsilon_v = \varepsilon_{xx} + \varepsilon_{xy}^2/\varepsilon_{yy}$ and $\mu_v = \mu_{xx} + \mu_{xy}^2/\mu_{yy}$ are Voigt relative permittivity and Voigt relative permeability, respectively.

Under such magnetization, the dispersion equation for the surface polaritons at the interface between vacuum and the given structure has the form [26]:

$$\kappa_1 g_v + \kappa_2 g_0 + ik_x g_0 \tilde{g}_{xy} \tilde{g}_{yy}^{-1} = 0,$$ **(16)**

where $\mu \to g$ for the TE mode, and $\varepsilon \to g$ for the TM mode, respectively.

We should note that in two particular cases of the gyroelectric and gyromagnetic superlattices, dispersion relation (16) coincides with Eq. (33) of Ref. [10] and Eq. (21) of Ref. [15], respectively, which verifies the obtained solution.

It follows from Fig. 3(b) that in the Voigt geometry the magnetic field vector in the TM mode has components $\{0, 0, H_z\}$ and it is parallel to the external magnetic field $\vec{M}$, which results in the absence of its interaction with the magnetic subsystem [33], [34]. Thus, hereinafter dispersion features only of the TE mode are of interest for which the dispersion equation for surface polaritons (16) can be rewritten as

$$\kappa_1 \mu_v + \kappa_2 \mu_0 + ik_x \mu_0 \mu_{xy} \mu_{yy}^{-1} = 0.$$ **(17)**

Importantly, since the dispersion equation consists of a term which is linearly depended on $k_x$ [the last term in (17)], the spectral characteristics of surface polaritons in the structure under study possess the nonreciprocal nature, i.e. $k_0(k_x) \neq k_0(-k_x)$.

A complete set of dispersion curves obtained from solution of Eq. (14) that outlines the passbands of the TE bulk polaritons as a function of the filling factor $\delta_m$ is presented in Fig. 4(a). One can see that behaviors of the TE bulk polaritons are quite trivial in the overwhelming majority of structure's configurations, namely there are two isolated passbands separated by a forbidden band. The upper passband is bounded laterally by the light line and its lower limit is restricted by the line at which $\mu_v = 0$. The bottom passband starts on the line where $\varepsilon_{zz} = 0$, and then approaches the asymptotic frequency ($k_x \to \infty$) at which $\mu_{xx} = 0$ [see, Eq. (14)].

In this study, we are mainly interested in those curves of the set which have greatly sloping branches and exhibit the closest approaching each other (i.e., they manifest the anti-crossing effect) or have a crossing point, since such dispersion behaviors correspond to the existence of the Morse critical points. Hereinafter the areas of interest in which these extreme states exist are denoted in figures by orange circles

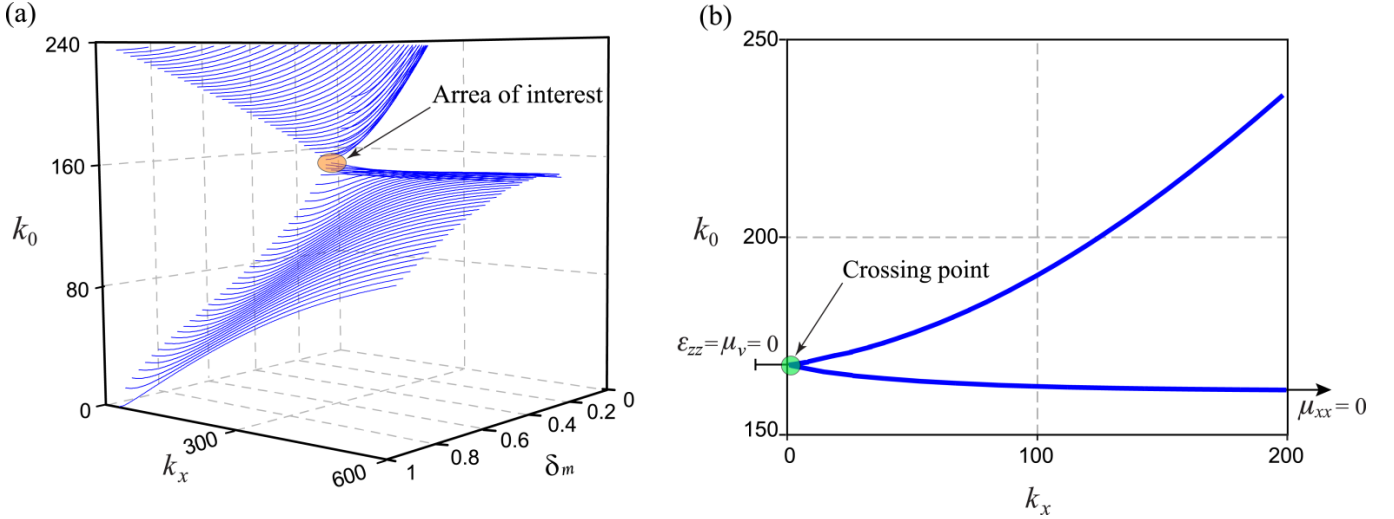

**Fig. 4.** (a) A set of dispersion curves of the TE bulk polaritons for different filling factor $\delta_m$ for the structure being in the Voigt geometry; (b) manifestation of the crossing effect in dispersion curves of the TE bulk polaritons at the particular value of filling factor $\delta_m = 0.132$. Parameters of magnetic constitutive layers are as in Fig. 2. For the semiconductor constitutive layers, parameters are: $f_p = \omega_p/2\pi = 10.5$ GHz, $f_c = \omega_c/2\pi = 9.5$ GHz, $\nu = 0$, $\varepsilon_l = 1.0$, $\mu_s = 1.0$.

It follows from Fig. 4(a) that in the Voigt configuration, both the anti-crossing ($\mathbb{H} < 0$) and crossing ($\mathbb{H} = 0$) effects with forming the bottom branch characterized by the anomalous dispersion (i.e., the strong mode interaction occurs) can be achieved in the composite structure with predominant impact of the semiconductor subsystem (further we stipulate that the composite system has predominant impact of either magnetic subsystem or semiconductor subsystem if $\delta_m \gg \delta_s$ or $\delta_s \gg \delta_m$, respectively). Moreover, since condition (B6) is met near the critical point, the bulk waves appear to be contradirectional.

The crossing effect found to be at $k_x \to 0$ (accidentally degenerate modes) for the particular configuration of the structure, when $\delta_m = 0.132$ [Fig. 4(b)]. Remarkably, such an extreme state corresponds to a particular frequency where $\varepsilon_{zz}$ and $\mu_v$ simultaneously acquire zero [26].

**B. Polar geometry**

In the polar geometry ($\vec{M} \parallel y$, $\vec{M} \perp \vec{k}$) the homogenized structure appears as a uniaxial crystal whose optical axis is directed along the $y$-axis [see, Fig. 3(a)]. In this case bulk polaritons split into two waves with field components $\{E_x, H_y, E_z\}$ and $\{H_x, E_y, H_z\}$, respectively [41], and their continua is outlined by two dispersion relations.

$$k_x^2 - k_0^2 \mu_{yy} \varepsilon_{yy} = 0, \quad (18)$$

$$k_x^2 - k_0^2 \mu_v \varepsilon_v (1 - \mu_{xz} \varepsilon_{xz} (\mu_{xx} \varepsilon_{xx})^{-1})^{-1} = 0, \quad (19)$$

where $\varepsilon_v = \varepsilon_{xx} + \varepsilon_{xz}^2/\varepsilon_{xx}$ and $\mu_v = \mu_{xz} + \mu_{xz}^2/\mu_{xx}$ are introduced as relative effective bulk permeability and permittivity, respectively

Hereinafter, we will distinguish these two kinds of waves as ordinary and extraordinary bulk polaritons, respectively (note, such a definition is common in the plasma physics [42]).

In order to elucidate the dispersion features of hybrid surface polaritons, the initial problem is decomposed into two particular solutions with respect to the vector $\vec{H}$ (EH-modes) and vector $\vec{E}$ (HE-modes) [10], [41]. In this way, the dispersion equation for surface polaritons is derived in the form:

$$\begin{aligned}
&\kappa_0^2 \tilde{g}_{xx} g_0^{-1}(\kappa_1^2 + \kappa_1 \kappa_2 + \kappa_2^2 - \varkappa_z^2) \\
&+ \kappa_0^2 \tilde{g}_{xx} g_0^{-1} \varkappa_y^2 \varsigma_{xz} \tilde{g}_{xz} (\varsigma_{yy} \tilde{g}_{xx})^{-1} \\
&+ \kappa_0 \left\{ (\kappa_1 + \kappa_2) \left[ \kappa_1 \kappa_2 + \varkappa_y^2 \tilde{g}_{xx} g_v (g_0 \varsigma_{yy})^{-1} \right] \right\} \\
&+ \tilde{g}_{xz} \varsigma_{xz}^{-1} \{ (\varkappa_z^4 - \varkappa_z^2 (\kappa_1 + \kappa_2) + \kappa_1^2 \kappa_2^2 ) \\
&+ \varkappa_y^2 (\varkappa_z^2 + \kappa_1 \kappa_2) \varsigma_{xz} \tilde{g}_{zz} (\varsigma_{yy} \tilde{g}_{xz})^{-1} \} = 0,
\end{aligned} \quad (20)$$

where $\varkappa_v^2 = k_x^2 - k_0^2 \varsigma_{vv}$, and two distinct substitutions $\varepsilon_{vv'} \to \tilde{g}_{vv'}$, $\varepsilon_0 \to g_0$, $\varepsilon_v \to g_v$ and $\varepsilon_{vv'} \to \tilde{g}_{vv'}$, $\mu_0 \to g_0$, $\mu_v \to g_v$ correspond to the problem resolving with respect to the vector $\vec{E}$ and vector $\vec{H}$, respectively; $\varepsilon_v$ and $\mu_v$ are the same as in Eqs. (18) and (19).

For two particular cases of the gyroelectric and gyromagnetic superlattices, dispersion relation (20) coincides with Eq. (23) of Ref. [10] and Eq. (21) of Ref. [13], respectively, which verifies the obtained solution.

Complete sets of dispersion curves calculated from the solution of equations (18) and (19) that outline the passbands of both ordinary (blue curves) and extraordinary (red curves) bulk polaritons as functions of the filling factor $\delta_m$ are presented in Fig. 5(a, b). Moreover, in order to discuss the observed crossing and anti-crossing effects more clearly, dispersion curves of both ordinary and extraordinary bulk polaritons are plotted in the $k_0 - k_x$ plane at the particular values of filling factor $\delta_m$ as shown in Fig. 5(c-f).

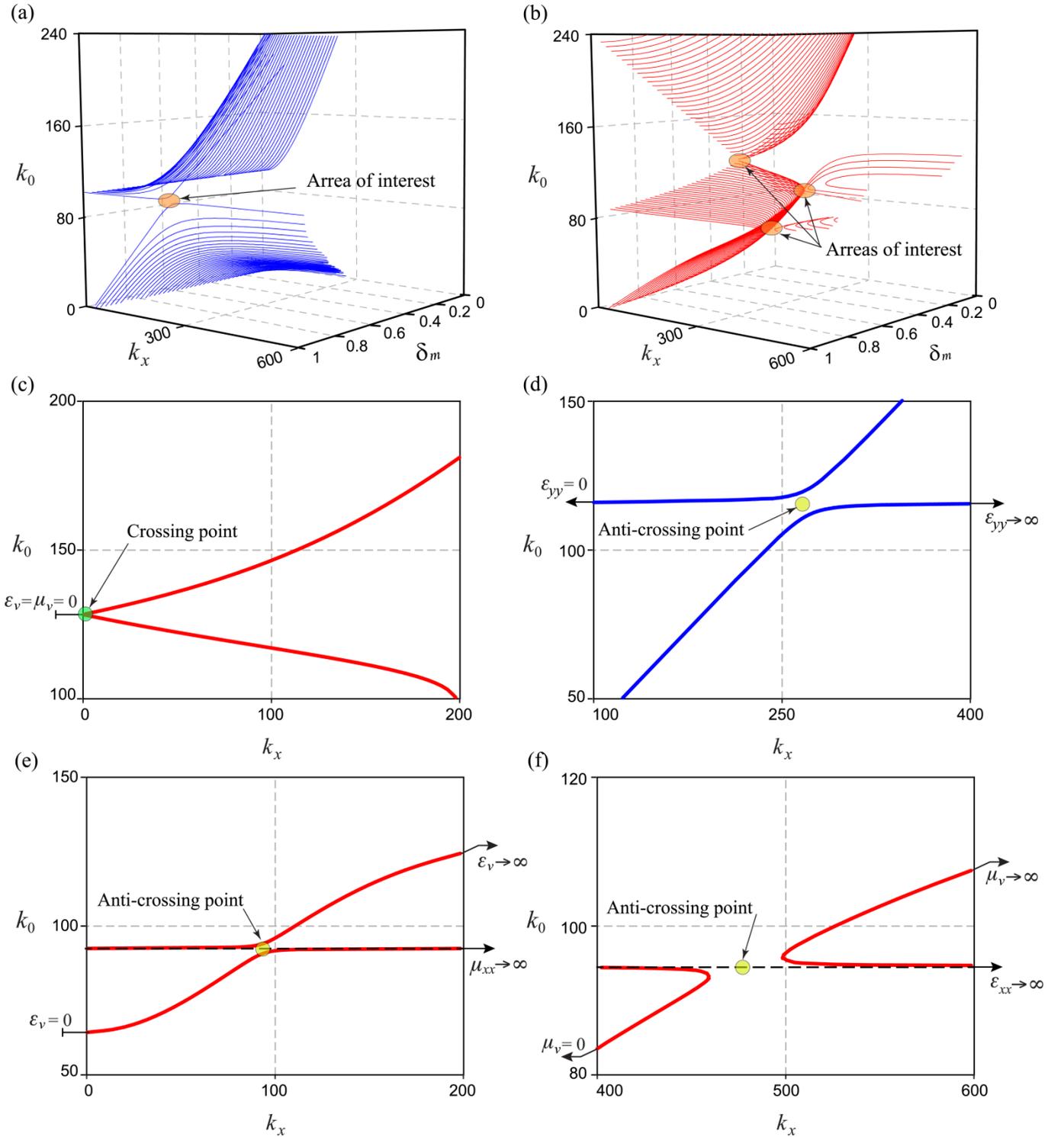

**Fig. 5.** Complete sets of dispersion curves of (a) ordinary (blue curves) and (b) extraordinary (red curves) bulk polaritons for different filling factor $\delta_m$ for the structure being in the polar geometry. Manifestation of (c) crossing and (d-f) anti-crossing effects in dispersion curves of bulk polaritons at the particular value of filling factor $\delta_m$; (c) $\delta_m = 0.267$; (d) $\delta_m = 0.99$; (e) $\delta_m = 0.05$; (f) $\delta_m = 0.99$. All structure constitutive parameters are as in Fig. 2.

From Fig. 5(a) one can conclude that there are two isolated areas of existence of ordinary bulk polaritons. The upper passband starts at the frequency where $\varepsilon_{yy} = 0$, while the bottom passband is bounded above by the asymptotic line where $\varepsilon_{yy} \to \infty$. Remarkably, the anti-crossing effect ($\mathbb{H} < 0$) between dispersion curves which restrict upper and bottom passbands of ordinary bulk polaritons can be observed in the structure with predominant impact of the magnetic subsystem (i.e., $\delta_m \gg \delta_s$) as presented in Fig. 5(a, d). Coupled modes exhibit an intermediate interaction for which an appearance of the flattened branches in dispersion curves is peculiar [see, for instance, Fig. B1(b)]. Besides, condition (B4) is met, so these waves are codirectional forward in the area nearly the critical point.

As show in Fig. 5(a) there are two separated areas of existence of extraordinary bulk polaritons for each particular filling factor $\delta_m$. Moreover, there are two possible combinations of conditions for their passbands which depend strictly on the value of filling factor $\delta_m$ [41]. Therefore, the particular critical filling factor is denoted here as $\delta_c$ at which $\varepsilon_v = \mu_v = 0$ at $k_x = 0$, and it is considered as a separation point between these two combinations. Thus the overall range of values $\delta_m$ can be separated into two sub-ranges $0 \leq \delta_m < \delta_c$ and $\delta_c < \delta_m \leq 1$, respectively.

The value of the critical filling factor $\delta_c$ depends on the constitutive parameters of magnetic and semiconductor layers, and for the structure under study it is $\delta_c = 0.267$. Such critical state in the effective parameters of the superlattice leads to absence of the forbidden band between the upper and bottom passbands as shown in Fig. 5(c). Remarkably, these dispersion branches appear to be coupled exactly at the frequency where $\varepsilon_v = \mu_v = 0$, i.e., the crossing effect ($\mathbb{H} = 0$) is found to be at $k_x \to 0$. Besides, such coupled waves are contradirectional, and they are described by condition (B6).

At the same time, for all present values of filling factor $\delta_m$ from the first sub-range $0 \leq \delta_m < \delta_c$, the lower limits of both the upper and bottom passbands are restricted by the lines at which $\varepsilon_v = 0$, as well as the bottom passband is bounded above by the asymptotic line where $\varepsilon_v \to \infty$. The upper passband is outlined by the set of following conditions [41]: $\varepsilon_v > 0, \mu_v > 0, \varepsilon_{xz}\mu_{xz}/\varepsilon_{xx}\mu_{xx} < 1$. The bottom passband exists when the set of conditions holds [41]: $\varepsilon_v > 0, \mu_v < 0$ and $\varepsilon_{xz}\mu_{xz}/\varepsilon_{xx}\mu_{xx} > 1$ or $\varepsilon_v > 0, \mu_v > 0$ and $\varepsilon_{xz}\mu_{xz}/\varepsilon_{xx}\mu_{xx} < 1$.

For all other values of filling factor $\delta_m$ (i.e., $\delta_c < \delta_m \leq 1$) the conditions for the upper passband are the same (i.e., $\varepsilon_v > 0$, $\mu_v > 0$, $\varepsilon_{xz}\mu_{xz}/\varepsilon_{xx}\mu_{xx} < 1$), whereas its lower limit is at the line where $\mu_v = 0$. The bottom passband exists when either the set of conditions $\varepsilon_v < 0$, $\mu_v > 0$ and $\varepsilon_{xz}\mu_{xz}/\varepsilon_{xx}\mu_{xx} > 1$ or $\varepsilon_v > 0$, $\mu_v > 0$ and $\varepsilon_{xz}\mu_{xz}/\varepsilon_{xx}\mu_{xx} < 1$ is satisfied. For this band the lower and upper limits are restricted by the lines at which $\mu_v = 0$ and $\mu_v \to \infty$, respectively.

From Fig. 5(a, b) one can conclude that the upper passband possesses typical behaviors in the $k_0 - k_x$ plane whereas the width and position of the bottom passbands are defined by the corresponding resonant frequencies of effective bulk permeability $\mu_v$ and effective bulk permittivity $\varepsilon_v$. In fact, these constitutive parameters are multipliers of the numerator of Eq. (19), whereas the denominator of this equation originates a singularity in the asymptotic line where $1 - \varepsilon_{xz}\mu_{xz}/\varepsilon_{xx}\mu_{xx} \to \infty$. Obviously, it corresponds to the cases when either $\mu_{xx} \to \infty$ or $\varepsilon_{xx} \to \infty$. This asymptotic line splits the bottom passbands into two separated sub-passbands which corresponds to the structures with predominant impact of semiconductor and magnetic subsystems, respectively [Fig. 5(e, f)]. Note, the intermediate interaction with forming flattened region in dispersion curves between bulk modes from these two sub-bands is observed nearly the Morse critical points ($\mathbb{H} < 0$). Such coupled waves are contradirectional since condition (B6) is met.

### C. Faraday geometry

In the Faraday geometry ($\vec{M} \parallel z$, $\vec{M} \parallel \vec{k}$) the superlattice is characterized by the biaxial anisotropy, for which, as already mentioned, the first anisotropy axis is associated with the structure periodicity (so it is directed along the $y$-axis) while the second anisotropy axis is a result of the external static magnetic field influence (so it is directed along the $z$-axis). For this geometry the bulk polaritons in the structure under study appear as right-handed and left-handed elliptically polarized waves [10], [12], [43] and their passbands are outlined by curves governed by the following dispersion law [43]:

$$\varkappa_x^2 \varkappa_y^2 - k_0^4 \varsigma_{xy} \varsigma_{yx} = 0, \tag{21}$$

where $\varkappa_v^2 = k_z^2 - k_0^2 \varsigma_{vv}$.

The surface polaritons are hybrid EH and HE-waves in the Faraday geometry and their dispersion relation can be written in the form [44]:

$$\begin{aligned}
&(\kappa_2 + \kappa_0 \tilde{g}_{zz})(\kappa_2^2 \varsigma_{yy} - \varkappa_y^2 \varsigma_{zz})\{\kappa_1 \zeta (\kappa_0^2 - k_z^2) \\
&\quad + \kappa_0 [k_1^2 \tilde{g}_{xy} \varsigma_{yy} - \varsigma_{zz}(k_0^2 \xi + k_z^2 \tilde{g}_{xy})]\} \\
&- (\kappa_1 + \kappa_0 \tilde{g}_{zz})(\kappa_1^2 \varsigma_{yy} - \varkappa_y^2 \varsigma_{zz})\{\kappa_2 \zeta (\kappa_0^2 - k_z^2) \\
&\quad + \kappa_0 [k_2^2 \tilde{g}_{xy} \varsigma_{yy} - \varsigma_{zz}(k_0^2 \xi + k_z^2 \tilde{g}_{xy})]\} = 0.
\end{aligned} \tag{22}$$

In Eq. (22), two distinct substitutions $\varepsilon_{vv'} \to \tilde{g}_{vv'}$, $\varepsilon_v \to g_v$ and $\mu_{vv'} \to \tilde{g}_{vv'}$, $\mu_v \to g_v$ correspond to the problem resolving with respect to the vector $\vec{E}$ and vector $\vec{H}$, respectively; $\varepsilon_v = \varepsilon_{xx} + \varepsilon_{xy}^2/\varepsilon_{yy}$; $\mu_v = \mu_{xx} + \mu_{xy}^2/\mu_{yy}$; $\zeta = g_v \tilde{g}_{yy} \varsigma_{yx}$; $\xi = \tilde{g}_{yy} \varsigma_{yx} - \tilde{g}_{xy} \varsigma_{yy}$; and the constant $g_0 = 1$ is omitted.

For the semiconductor superlattice, the dispersion equation (22) agrees with Eq. (38) of Ref. [10], while in the case of magnetic superlattice it coincides with Eq. (13) of Ref. [12], which verifies the obtained solution.

Complete sets of dispersion curves that outline the bands of existence of the bulk polaritons [see, Eq. (21)] as functions of filling factor $\delta_m$ are presented in Fig. 6(a, b) for right-handed (blue curves) and left-handed (red curves) elliptical polarizations. From these figures one can conclude that there is a pair of corresponding sets of dispersion curves separated by a forbidden band for bulk polaritons of each polarization. The dispersion curves of right-handed elliptically polarized bulk waves demonstrate quite trivial behaviors and they completely inherit characteristics of right-handed circularly polarized waves of the corresponding reference semiconductor or magnetic medium (see, for instant, [43]). Contrariwise, the dispersion features of left-handed elliptically polarized waves are much more complicated being strongly dependent on filling factor $\delta_m$ and resonant frequencies of constitutive parameters of both semiconductor and magnetic underlying materials. Therefore, in what follows we are interested only in the consideration of dispersion features of bulk polaritons having left-handed polarization.

One can conclude that the dispersion characteristics of left-handed elliptically polarized bulk waves of the given gyroelectromagnetic structure are different from those related to convenient gyroelectric and gyromagnetic media. Indeed, in contrast to the characteristics of left-handed circularly polarized waves of the corresponding reference media whose passband has no discontinuity, the passband of left-handed elliptically polarized bulk polaritons of the superlattice is separated into two distinct areas. This separation appears nearly the frequency at which the resonances of the functions $\varepsilon_{xy}$ are $\mu_{xy}$ occur for the structures with predominant impact of the magnetic and semiconductor subsystems, respectively. Also, we should note that the exchange by the critical conditions for the asymptotic lines (at which $k_z \to \infty$) between the bottom passbands of left-handed and right-handed elliptically polarized bulk waves appears at the particular frequency where $\varepsilon_{xy}$ and $\mu_{xy}$ simultaneously tend to infinity.

The dispersion curves of bulk waves which have left-handed polarization demonstrate significant variation of their slope having subsequent branches with normal and anomalous dispersion that possess approaching at some points (extreme states) as depicted in Fig. 6(c, e).

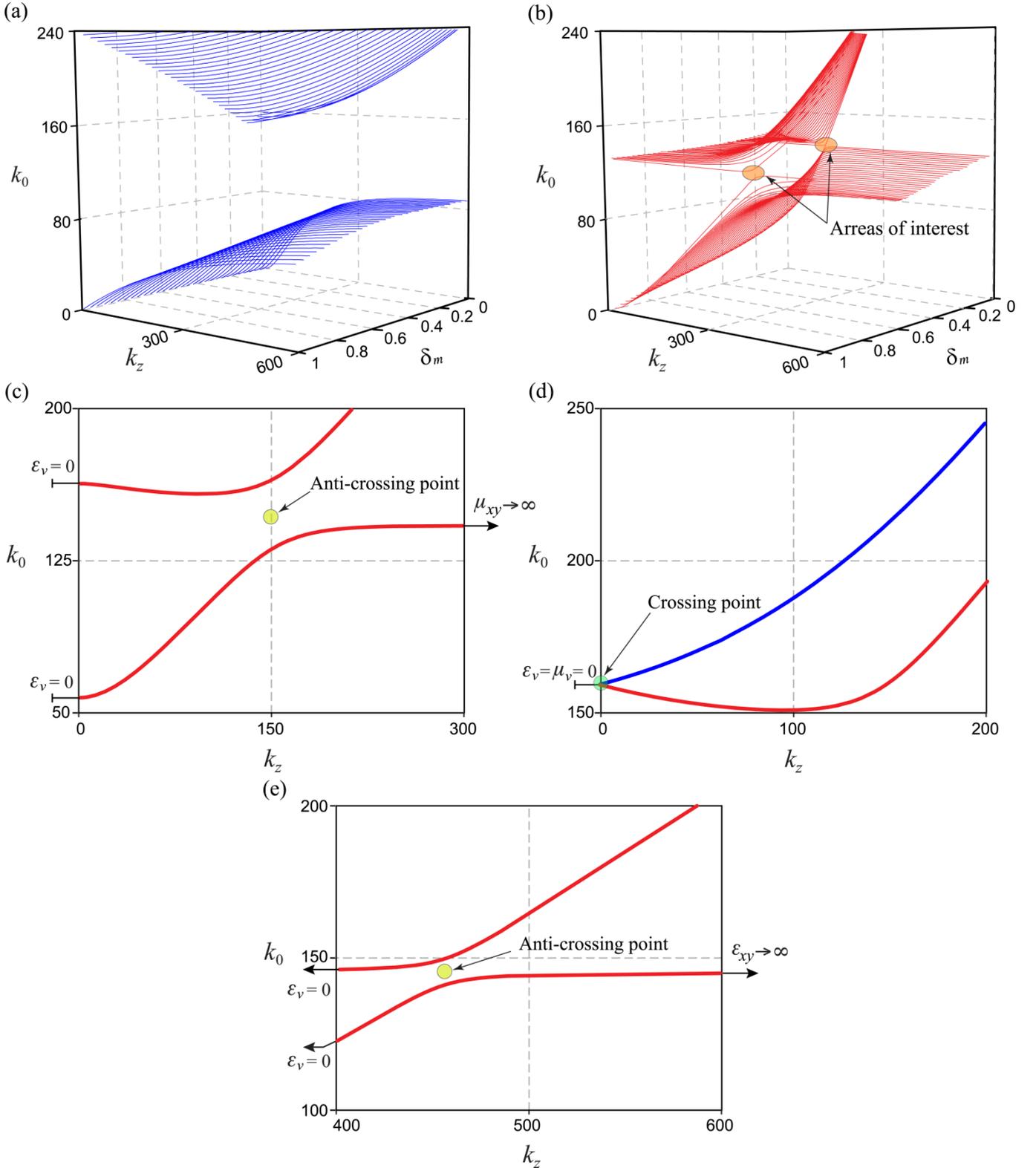

**Fig. 6.** (a, b) Complete sets of dispersion curves of both right-handed (blue curves) and left-handed (red curves) elliptically polarized bulk polaritons for different filling factor $\delta_m$ for the structure being in the Faraday geometry. Manifestation of (c, e) anti-crossing and (d) crossing effects in dispersion curves of bulk polaritons at the particular value of filling factor $\delta_m$; (c) $\delta_m = 0.15$; (d) $\delta_m = 0.1$; (e) $\delta_m = 0.95$. All structure constitutive parameters are as in Fig. 2.

The features of these dispersion curves in the vicinity of the critical points are different for the structure which has predominant impact either semiconductor (i.e., $\delta_s \gg \delta_m$) or magnetic (i.e., $\delta_m \gg \delta_s$) subsystem. The corresponding modes are contradirectional (i.e., condition (B6) is met) and they demonstrate strong interaction nearly the Morse critical point (the anti-crossing effect with $\mathbb{H} < 0$) in the case when $\delta_s \gg \delta_m$. The modes acquire an intermediate interaction in the superlattice with predominant impact of the magnetic subsystem as depicted in Fig. 6(e).

Moreover, a particular extreme state in dispersion curves is found out where the upper branches of left-handed and right-handed elliptically polarized bulk waves merge with each other [see, Fig. 6(d)], and crossing point ($\mathbb{H} = 0$) occurs at $k_z \to 0$. Likewise to the polar geometry, such extreme state corresponds to the particular superlattice configuration where $\varepsilon_v = \mu_v = 0$ at $k_z = 0$. So, the interacting waves are degenerated at this point and possess the contradirectional propagation.

We should note that in [44] some unusual and counter-intuitive consequences of such behaviors of the dispersion curves (e.g., backward waves propagation, reversed Doppler shift, reversed Cherenkov radiation, atypical singularities in the density of states, etc.) for the TE and TM modes of an axially uniform waveguide are discussed, and it is emphasized that these effects are of considerable significance for practical applications.

### D. Surface polaritons

Finally, in order to obtain the dispersion curves $k_0(k)$ of surface polaritons in both the polar and Faraday geometries, characteristic equations (20) and (22) are solved numerically [41], [43]. Here as previously, the problem is decomposed into two particular solutions with respect to the vector $\vec{H}$ (EH waves) and vector $\vec{E}$ (HE waves) [10], [12]. Note, as evident from equations (20) and (22) $k_x$ and $k_z$ appear only in even powers, so the dispersion curves of the surface waves for these geometries appear to be reciprocal.

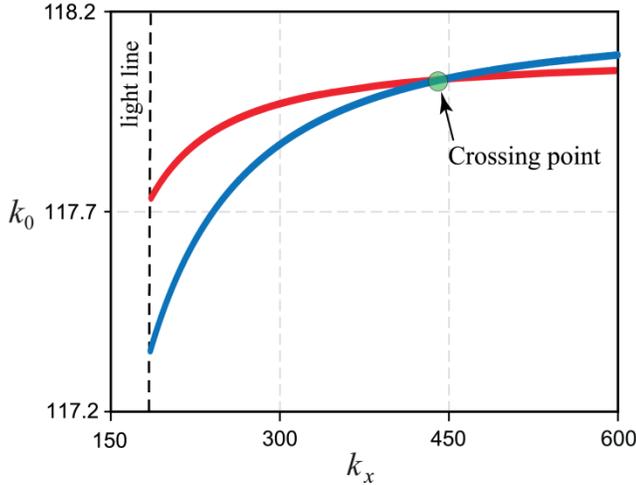
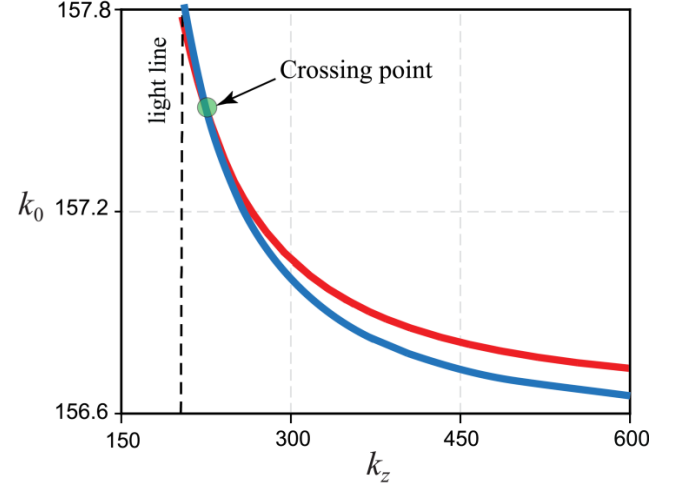

**Fig. 7.** Manifestation of the crossing effect in dispersion curves of hybrid EH (red line) and HE (blue line) surface polaritons; (a) polar geometry ($\delta_m = 0.27$); (b) Faraday geometry ($\delta_m = 0.15$). All structure constitutive parameters are as in Fig. 2.

For both problem considerations, dispersion equations (20) and (22) have four roots from which two physically correct ones (denoted here as $\kappa_1$ and $\kappa_2$) are selected [10] ensuring they correspond to the attenuating waves. In papers [10] and [12] it was noted that depending on the position in the $k_0$–$k$ plane, in non-dissipative systems the following combinations between two roots $\kappa_1$ and $\kappa_2$ may arise: (i) both roots are real and positive (bona fide surface modes); (ii) one root is real and the other is pure imaginary, or vice versa (pseudosurface modes); (iii) both roots are complex in which case they are conjugate (generalized surface modes); and (iv) both roots are pure imaginary (the propagation is forbidden). In our study we consider only bona fide surface mode. Nevertheless, we should note that in both the polar and Faraday geometries pseudo-surface waves (which attenuate only on one side of the surface) can also be supported [10], [45].

Among all possible appearance of dispersion curves of the surface polaritons we are only interested in those ones which manifest the crossing or anti-crossing effect. The search of their existence implies solving an optimization problem, where for the crossing effect the degeneracy point should be found. For the anti-crossing effect the critical points are defined from calculation of the first $\mathfrak{D}'(k, k_0)$ and second $\mathfrak{D}''(k, k_0)$ partial derivatives of equation (20) or (22) with respect to $k$ and $k_0$ for both EH and HE modes. During the solution of this problem the period and constitutive parameters of the underlying materials of the superlattice are fixed, and the search for the effects manifestation is preceded via altering the filling factors $\delta_m$ and $\delta_s$ within the period. The found crossing points are depicted in Fig. 8(a, b) by green circles. We should note that in both configurations the degeneracy points for the EH (red line) and HE (blue line) surface polaritons can be obtained only for the case of composite structure with a predominant impact of the semiconductor subsystem. In particular, this state is found out to be at the values of filling factor $\delta_m = 0.27$ and $\delta_m = 0.15$ for the polar and Faraday geometries, respectively. Remarkable, in the Faraday geometry both dispersion curves possess an anomalous dispersion line, namely they start on the light line and fall just to the right of the light line, and then they flatten out and approach an asymptotic limit for large values of $k$. Contrariwise to the Faraday geometry, in the polar geometry both dispersion curves demonstrate normal dispersion.

## 4. CONCLUSIONS

To conclude, in this paper we have studied dispersion features of both bulk and surface polaritons in a magnetic-semiconductor superlattice influenced by an external static magnetic field. The investigation was carried out under an assumption that all characteristic dimensions of the given superlattice satisfy the long-wavelength limit, thus the homogenization procedures from the effective medium theory was involved, and the superlattice was represented as a gyroelectromagnetic uniform medium characterized by the tensors of effective permeability and effective permittivity.

The general theory of polaritons in the gyroelectromagnetic medium whose permittivity as well as permeability simultaneously are tensor quantities was developed. Three particular cases of the magnetization, namely the Voigt, polar and Faraday geometries were discussed in detail.

The crossing and anti-crossing effects in the dispersion curves of both surface and bulk polaritons have been identified and investigated with an assistance of the analytical theory about the Morse critical points.

We argue that the discussed dispersion features of polaritons identified in the magnetic-semiconductor superlattice under study have a fundamental nature and are common to different types of waves and waveguide systems

## APPENDIX A: EFFECTIVE CONSTITUTIVE PARAMETERS OF A SUPERLATTICE

In order to obtain expressions for the tensors of effective permeability and effective permittivity of the superlattice in a general form, constitutive equations, $\vec{B} = \mu \vec{H}$ and $\vec{D} = \varepsilon \vec{E}$ for magnetic ($0 < z < d_m$) and semiconductor ($d_m < z < L$) layers are represented as follow [26]:

$$Q_\nu^{(j)} = \sum_{\nu'} g_{\nu\nu'}^{(j)} P_{\nu'}^{(j)}, \tag{A1}$$

where $\vec{Q}$ takes values of the magnetic and electric flux densities $\vec{B}$ and $\vec{D}$, respectively; $\vec{P}$ is varied between the magnetic and electric field strengths $\vec{H}$ and $\vec{E}$; $g$ is substituted for permeability and permittivity $\mu$ and $\varepsilon$; the superscript $j$ is introduced to distinguish between magnetic ($m \to j$) and semiconductor ($s \to j$) layers, and $\nu\nu'$ iterates over $x$, $y$ and $z$.

In the given structure geometry, the interfaces between adjacent layers within the superlattice lie in the $x - z$ plane and they are normal to the $y$-axis, thus the field components $P_x^{(j)}$, $P_z^{(j)}$ and $Q_y^{(j)}$ are continuous at the interfaces. Thus, the normal component $P_y^{(j)}$ can be expressed from (A1) in terms of the continuous components of the field as follow:

$$P_y^{(j)} = -\frac{g_{yx}^{(j)}}{g_{yy}^{(j)}} P_x^{(j)} + \frac{1}{g_{yy}^{(j)}} Q_y^{(j)} - \frac{g_{yz}^{(j)}}{g_{yy}^{(j)}} P_z^{(j)}, \tag{A2}$$

and substituted into equations for components $Q_x^{(j)}$ and $Q_z^{(j)}$:

$$Q_x^{(j)} = \left(g_{xx}^{(j)} - \frac{g_{xy}^{(j)} g_{yx}^{(j)}}{g_{yy}^{(j)}}\right) P_x^{(j)} + \frac{g_{xy}^{(j)}}{g_{yy}^{(j)}} Q_y^{(j)} + \left(g_{xz}^{(j)} - \frac{g_{xy}^{(j)} g_{yz}^{(j)}}{g_{yy}^{(j)}}\right) P_z^{(j)},$$

$$Q_z^{(j)} = \left(g_{zx}^{(j)} - \frac{g_{zy}^{(j)} g_{yx}^{(j)}}{g_{yy}^{(j)}}\right) P_x^{(j)} + \frac{g_{zy}^{(j)}}{g_{yy}^{(j)}} Q_y^{(j)} + \left(g_{zz}^{(j)} - \frac{g_{zy}^{(j)} g_{yz}^{(j)}}{g_{yy}^{(j)}}\right) P_z^{(j)}. \tag{A3}$$

Relations (A2) and (A3) are then used for the fields averaging [24].

With taking into account the long-wavelength limit, the fields $\vec{P}^{(j)}$ and $\vec{Q}^{(j)}$ inside the layers are considered to be constant and the *averaged* (Maxwell) fields $\langle\vec{Q}\rangle$ and $\langle\vec{P}\rangle$ can be determined by the equalities

$$\langle \vec{P} \rangle = \frac{1}{L} \sum_j \vec{P}^{(j)} d_j, \quad \langle \vec{Q} \rangle = \frac{1}{L} \sum_j \vec{Q}^{(j)} d_j, \tag{A4}$$

In view of the continuity of components $P_x^{(j)}$, $P_z^{(j)}$ and $Q_y^{(j)}$, it follows that

$$\langle P_x \rangle = P_x^{(j)}, \quad \langle P_z \rangle = P_z^{(j)}, \quad \langle Q_y \rangle = Q_y^{(j)}, \tag{A5}$$

and with using Eqs. (A2) and (A3), we can obtain the relations between the averaged fields components in the next form:

$$\langle Q_x \rangle = \alpha_{xx} \langle P_x \rangle + \gamma_{xy} \langle Q_y \rangle + \alpha_{xz} \langle P_z \rangle,$$
$$\langle P_y \rangle = \beta_{yx} \langle P_x \rangle + \beta_{yy} \langle Q_y \rangle + \beta_{yz} \langle P_z \rangle, \tag{A6}$$
$$\langle Q_z \rangle = \alpha_{zz} \langle P_x \rangle + \gamma_{zy} \langle Q_y \rangle + \alpha_{zz} \langle P_z \rangle,$$

here we used following designations $\alpha_{\nu\nu'} = \sum_j \left(g_{\nu\nu'}^{(j)} - g_{\nu y}^{(j)} g_{y\nu'}^{(j)}/g_{yy}^{(j)}\right) \delta_j$, $\beta_{yy} = \sum_j \left(1/g_{yy}^{(j)}\right) \delta_j$, $\beta_{yy} = \sum_j \left(g_{y\nu'}^{(j)}/g_{yy}^{(j)}\right) \delta_j$, $\gamma_{\nu y} = \sum_j \left(g_{\nu y}^{(j)}/g_{yy}^{(j)}\right) \delta_j$, $\delta_j = d_j/L$ is filling factor, and $\nu\nu'$ iterates over $x$, $z$.

Expressing $\langle Q_y \rangle$ from the second equation in system (A6) and substituting it into the rest two equations, the constitutive equations for the flux densities of the effective medium $\langle \vec{Q} \rangle = \hat{g}_{eff} \langle \vec{P} \rangle$ can be derived, where $\hat{g}_{eff}$ is a tensor quantity

$$\hat{g}_{eff} = \begin{pmatrix} \tilde{\alpha}_{xx} & \tilde{\gamma}_{xy} & \tilde{\alpha}_{xz} \\ \tilde{\beta}_{yx} & \tilde{\beta}_{yy} & \tilde{\beta}_{yz} \\ \tilde{\alpha}_{zx} & \tilde{\gamma}_{zy} & \tilde{\alpha}_{zz} \end{pmatrix} = \begin{pmatrix} \tilde{g}_{xx} & \tilde{g}_{xy} & \tilde{g}_{xz} \\ \tilde{g}_{yx} & \tilde{g}_{yy} & \tilde{g}_{yz} \\ \tilde{g}_{zx} & \tilde{g}_{zy} & \tilde{g}_{zz} \end{pmatrix}, \tag{A7}$$

with components $\tilde{\alpha}_{\nu\nu'} = \alpha_{\nu\nu'} - \beta_{y\nu'} \gamma_{\nu y}/\beta_{yy}$, $\tilde{\beta}_{yy} = 1/\tilde{\beta}_{yy}$, $\tilde{\beta}_{y\nu'} = -\beta_{y\nu'}/\beta_{yy}$, and $\tilde{\gamma}_{\nu y} = -\gamma_{\nu y}/\beta_{yy}$.

The expressions for tensors components of the underlying constitutive parameters of magnetic ($\hat{\mu}_m \to \hat{g}^{(m)}$) and semiconductor ($\hat{\varepsilon}_s \to \hat{g}^{(s)}$) layers depend on the orientation of the external magnetic field $\vec{M}$ in the $y - z$ plane which is defined by the angle $\theta$ in the form

$$\hat{g}^{(j)} = \begin{pmatrix} g_1 & i\zeta g_2 & i\xi g_2 \\ -i\zeta g_2 & \zeta^2 g_1 + \xi^2 g_3 & \zeta\xi(g_1 - g_3) \\ -i\xi g_2 & \zeta\xi(g_1 - g_3) & \zeta^2 g_3 + \xi^2 g_1 \end{pmatrix}, \tag{A8}$$

where $\zeta = \sin\theta$ and $\xi = \cos\theta$.

For magnetic layers [29] the components of tensor $\hat{g}^{(m)}$ are: $g_1 = 1 + \chi' + i\chi''$, $g_2 = \Omega' + i\Omega''$, $g_3 = 1$ and $\chi' = \omega_0 \omega_m [\omega_0^2 - \omega^2(1 - b^2)]D^{-1}$, $\chi'' = \omega \omega_m b[\omega_0^2 + \omega^2(1 + b^2)]D^{-1}$, $\Omega' = \omega \omega_m [\omega_0^2 - \omega^2(1 + b^2)]D^{-1}$, $\Omega'' = 2\omega^2 \omega_0 \omega_m b D^{-1}$, $D = [\omega_0^2 - \omega^2(1 + b^2)]^2 + 4\omega_0^2 \omega b^2$, where $\omega_0$ is the Larmor frequency and b is a dimensionless damping constant.

For semiconductor layers [23] the components of tensor $\hat{g}^{(s)}$ are: $g_1 = \varepsilon_l[1 - \omega_p^2(\omega + i\nu)[\omega((\omega + i\nu)^2 - \omega_c^2)]^{-1}]$, $g_2 = \varepsilon_l \omega_p^2 \omega_c [\omega((\omega + i\nu)^2 - \omega_c^2)]^{-1}$, $g_3 = \varepsilon_l \left[1 - \left[\omega_p^2(\omega + i\nu)\right]^{-1}\right]$, where $\varepsilon_l$ is the part of permittivity attributed to the lattice, $\omega_p$ is the plasma frequency, $\omega_c$ is the cyclotron frequency and $\nu$ is the electron collision frequency in plasma.

Permittivity $\varepsilon_m$ of the magnetic layers as well as permeability $\mu_s$ of the semiconductor layers are scalar quantities.

Hereinafter, we consider two specific orientations of the external magnetic field vector $\vec{M}$ with respect to the superlattice's interface (see, Fig. 1), namely: (i) the polar configuration in which $\theta = 0$ and the vector $\vec{M}$ is parallel to the surface normal ($\vec{M} \parallel y$); (ii) $\theta = \pi/2$ and the vector $\vec{M}$ is parallel to the surface plane ($\vec{M} \parallel z$), which is inherent in both the Voigt and Faraday configurations.

When $\vec{M} \parallel y$ then $\zeta = 0, \xi = 1$, and tensor (A8) is reduced to the form

$$\hat{g}^{(j)} = \begin{pmatrix} g_1 & 0 & ig_2 \\ 0 & g_3 & 0 \\ -ig_2 & 0 & g_1 \end{pmatrix}, \tag{A9}$$

and for the components of tensor (A7) we have following expressions: $\tilde{\gamma}_{xy} = \tilde{\gamma}_{zy} = \tilde{\beta}_{yx} = \tilde{\beta}_{yz} = 0$, $\tilde{\alpha}_{xx} = g_{xx}^{(m)} \delta_m + g_{xx}^{(s)} \delta_s$, $\tilde{\alpha}_{zz} = g_{zz}^{(m)} \delta_m + g_{zz}^{(s)} \delta_s$, $\tilde{\alpha}_{xz} = -\tilde{\alpha}_{zx} = g_{zx}^{(m)} \delta_m + g_{zx}^{(s)} \delta_s$, $\tilde{\beta}_{yy} = g_{yy}^{(m)} g_{yy}^{(s)} \tau$, where $\tau = \left(g_{yy}^{(m)} \delta_s + g_{yy}^{(s)} \delta_m\right)^{-1}$.

In the second case, when $\vec{M} \parallel z$ then $\zeta = 1, \xi = 0$, and tensor (A8) has the form

$$\hat{g}^{(j)} = \begin{pmatrix} g_1 & ig_2 & 0 \\ -ig_2 & g_1 & 0 \\ 0 & 0 & g_3 \end{pmatrix}. \quad \textbf{(A10)}$$

In this configuration the components of tensor (A7) can be written as follows: $\tilde{\alpha}_{xz} = \tilde{\alpha}_{zx} = \tilde{\gamma}_{zy} = \tilde{\beta}_{yz} = 0$, $\tilde{\alpha}_{xx} = g_{xx}^{(m)}\delta_m + g_{xx}^{(s)}\delta_s + \left(g_{xy}^{(m)} - g_{xy}^{(s)}\right)^2 \delta_m \delta_s \tau$, $\tilde{\alpha}_{zz} = g_{zz}^{(m)}\delta_m + g_{zz}^{(s)}\delta_s$, $\tilde{\gamma}_{xy} = -\tilde{\beta}_{yx} = \left(g_{xy}^{(m)} g_{yy}^{(s)}\delta_m + g_{xy}^{(s)} g_{yy}^{(m)}\delta_s\right)\tau$, $\tilde{\beta}_{yy} = g_{yy}^{(m)} g_{yy}^{(s)}\tau$.

## APPENDIX B: THEORY OF MORSE CRITICAL POINTS. MODE COUPLING PHENOMENA

For brevity, obtained dispersion equations for bulk and surface polaritons are denoted in the form:

$$\mathfrak{D}(k, k_0) = 0, \quad \textbf{(B1)}$$

Numerical solution of equation (B1), gives a set of dispersion curves $k_0(k)$ of polaritons which can contain both regular and singular (critical) points. The regular points draw dispersion curves of a classical form possessing either normal or anomalous dispersion line, at which a small variation in $k_0$ results in a smooth changing in the form of the curves. Besides, some situations are possible when a slight variation in $k_0$ leads to a very sharp (catastrophic) changing in the form of dispersion curves. Such singularities (extreme states) can be accompanied by mutual coupling phenomena of modes which are further of our interest.

From the mathematical point of view the found extreme states in dispersion curves exist in the region where the differential $\mathfrak{D}'(k, k_0)$ of dispersion equation (B1) vanishes (see, for example, Fig. B1). These extreme states can be carefully identified and studied involving the approach based on the theory of the Morse critical points from the catastrophe theory [35]. This treatment has been originally applied for study of open waveguides and resonators [36], [37] and later it has been extended to more complex waveguide structures [38]–[40]. From viewpoint of this theory the presence of the Morse critical points are generally defined by a set of nonlinear differential equations written in the form [38]:

$$\begin{aligned} \mathfrak{D}'_k(k, k_0)|_{(k^m, k_0^m)} &= \mathfrak{D}'_{k_0}(k, k_0)|_{(k^m, k_0^m)} = 0, \\ \mathbb{H} &= [h_{11}h_{22} - h_{12}h_{21}]|_{(k^m, k_0^m)} \neq 0, \end{aligned} \quad \textbf{(B2)}$$

where $(k^m, k_0^m)$ are coordinates in the $k - k_0$ plane of a particular $m$-th Morse critical point, the subscripts $k$ and $k_0$ near the letter $\mathfrak{D}$ define corresponding partial derivatives $\partial/\partial k$ and $\partial/\partial k_0$, and $\mathbb{H}$ is the Hessian determinant with elements:

$$\begin{aligned} h_{11} &= \mathfrak{D}''_{kk}(k, k_0), & h_{12} &= \mathfrak{D}''_{kk_0}(k, k_0), \\ h_{21} &= \mathfrak{D}''_{k_0 k}(k, k_0), & h_{22} &= \mathfrak{D}''_{k_0 k_0}(k, k_0). \end{aligned} \quad \textbf{(B3)}$$

The type of each extreme state defined by set of equations (B2) can be uniquely identified from the sign of the Hessian determinant [38]. For instance, when $\mathbb{H} < 0$ the corresponding Morse critical point represents a saddle point which occurs in the region of a modal coupling (the *anti-crossing* effect), whereas in the case of degeneracy, when $\mathbb{H} = 0$, it is a non-isolated critical point (the *crossing* effect). In the case when $\mathbb{H} > 0$ the Morse critical point defines either a local minimum or maximum (this case is not considered here). In what follows we distinguish found critical points by green and yellow circles for the crossing and anti-crossing effects, respectively, as shown in Fig. B1.

In general, when conditions (B2) are met, the type of interacting modes in the vicinity of the corresponding Morse critical point can be defined as follows [38]:

*Codirectional forward:* $h_{12}/h_{11} < 0,\ h_{22}/h_{11} > 0;$ **(B4)**

*Codirectional backward:* $h_{12}/h_{11} > 0,\ h_{22}/h_{11} > 0;$ **(B5)**

*Contradirectional:* $h_{12}/h_{11} > 0,\ h_{22}/h_{11} < 0.$ **(B6)**

The modes interaction strength within the found extreme states in the region of their coupling can be identified considering the classification introduced in paper [44], which concerns on the modes behaviors appearing in axial waveguides. In particular, (i) a *weak interaction* takes place when frequency band gap between dispersion curves of interacting modes is high enough [Fig. B1(a)]; (ii) an *intermediate interaction* of modes leads to formation of very flattened parts in the dispersion curves [Fig. B1(b)]; (iii) a *strong interaction* appears when the repulsion between modes is strong enough resulting in formation of dispersion curve having anomalous dispersion line [Fig. B1(c)]; (iv) an *accidental degeneracy* arises when two dispersion branches are merged within the critical point which leads to nonzero group velocities at $k = 0$ [Fig. B1(d)].

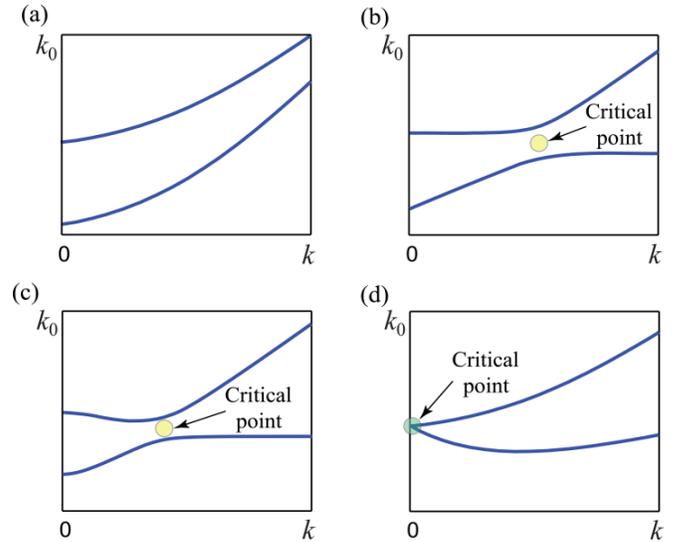

**Fig. B1.** Sketch of the band diagrams presenting different kinds of interaction between two neighboring modes: (a) weak interaction; (b) intermediate interaction; (c) strong interaction; (d) accidental degeneracy. Areas where the critical points exist are pointed by yellow and green circles.